\documentclass[thmsa]{article}
\usepackage{amssymb}
\usepackage{amsmath}
\usepackage{setspace}
\usepackage{graphicx}

\input tcilatex
\begin{document}

\title{Snapshot Models of Undocumented Immigration}
\author{Scott Rodilitz\thanks{%
Yale School of Management, 165 Whitney Avenue, New Haven 06511, CT. email:
scott.rodilitz@yale.edu} and Edward H. Kaplan\thanks{%
William N. and Marie A. Beach Professor of Operations Research, Professor of
Public Health, Professor of Engineering, Yale School of Management, 165
Whitney Avenue, New Haven 06511, CT. email: edward.kaplan@yale.edu}}
\date{(February 2020: Please do not cite without permission of authors)}
\maketitle

\begin{abstract}
The Mexican Migration Project (MMP) is a study that includes samples of
undocumented Mexican immigrants to the United States after their return to
Mexico. Of particular interest are the departure and return dates of a
sampled migrant's most recent sojourn in the United States, and the total
number of such journeys undertaken by that migrant household, for these data
enable the construction of data-driven undocumented immigration models.
However, such data are subject to an extreme physical bias, for to be
included in such a sample, a migrant must have returned to Mexico by the
time of the survey, excluding those undocumented immigrants still in the US.
In our analysis, we account for this bias by jointly modeling trip timing
and duration to produce the likelihood of observing the data in such a
"snapshot" sample. Our analysis characterizes undocumented migration flows
including single visit migrants, repeat visitors, and "retirement" from
circular migration. Starting with 1987, we apply our models to 30 annual
random snapshot surveys of returned undocumented Mexican migrants accounting
for undocumented Mexican migration from 1980--2016. Contrary to published
estimates based on these same data, our results imply migrants remain in the
US much longer than previously estimated based on analysis that ignored the
physical snapshot bias. Scaling to population quantities, we produce lower
bounds on the total number of undocumented immigrants that are much larger
than conventional estimates based on US-based census-linked surveys, and
broadly consistent with the estimates reported by Fazel-Zarandi, Feinstein
and Kaplan (2018).
\end{abstract}

\doublespacing

\section{Introduction and Motivation}

The number of undocumented immigrants that reside in the United States is
one of the most important and controversial quantities at the heart of the
US immigration debate. For many years, estimates produced by organizations
like the Pew Research Center in the neighborhood of 11 million undocumented
immigrants have persisted (Pew Research Center, 2018). These estimates are
deduced via the \textquotedblleft residual method,\textquotedblright\ which
relies on estimating the total number of US residents born outside of the
United States via extrapolation from questionnaires administered as part of
the American Community Survey (https://www.census.gov/programs-surveys/acs)
and the Current Population Survey
(https://www.census.gov/programs-surveys/cps.html), and subtracting out the
number of those born outside of the US who are in the country legally as
determined from government records such as visas, citizenship documents, etc.

Residual method estimates of the size of the undocumented population have
been questioned, however, due to the fact that persons in the United States
illegally have strong incentives not to cooperate with government surveys.
That the missing data rates for questions regarding place of birth are much
higher than for less troublesome questions such as age or gender lends
credibility to this concern (Kaplan, 2019). For example, in the 2017 version
of the American Community Survey, only 1.7\% of survey participants refused
to provide their age, in contrast with 9.3\% who refused to provide their
place of birth
(https://www.census.gov/acs/www/methodology/sample-size-and-data-quality/item-allocation-rates/). While statistical imputation methods are employed to adjust for such missing data, those methods presume that data are missing at random (Andridge and Little, 2010). When data regarding place of birth is sought from potentially undocumented immigrants, however, one must at least suspect that missing data are missing on purpose.

Rather than relying upon the residual method, Fazel-Zarandi, Feinstein and
Kaplan (2018) developed a mathematical model of the population of
undocumented immigrants from basic demographic principles: the population at
some point in time equals the initial population, plus the cumulation of all
population inflows minus all population outflows over time. The data for
estimating inflows and outflows was operational in nature, based largely on
data from the Department of Homeland Security (DHS) documenting border
apprehensions, visa overstays, and deportations, in addition to voluntary
emigration by immigrants out of the US at rates reported in the academic
literature on the subject. That model estimated conservatively that as of
the end of 2016 there were at least 16.7 million undocumented immigrants in
the US, while over one million simulations accounting for parameter
uncertainty, the mean population estimate was 22 million, essentially twice
the results based on the residual method.

These new estimates were themselves criticized for presumably
underestimating undocumented immigrants' voluntary emigration rates out of
the US. This criticism, argued by Capps \textit{et al} (2018), was driven by
new emigration estimates obtained from the Mexican Migration Project (Durand
and Massey, 2006). As will be discussed in greater detail later, the
relevant data collected by the Mexican Migration Project (MMP) consist of
sojourn times spent in the United States by former Mexican undocumented
immigrants \textit{who were sampled in Mexico after their return}. These
data thus comprise a \textit{snapshot sample} (Kaplan, 1997), and as is well
known such samples are subject to an extreme physical bias. In the case of
the MMP data, a person could only be included in the sample after returning
to Mexico, ruling out all undocumented immigrants remaining in the United
States at the time sampling took place. This bias in turn leads to severe 
\textit{underestimation} of the duration migrants spend in the United
States, for undocumented immigrants with long sojourn times are much less
likely to have returned to Mexico in time for the survey! This snapshot bias
easily explains why Capps \textit{et al} (2018) estimated much shorter
sojourn times (and hence much larger emigration rates) than those considered
by Fazel-Zarandi, Feinstein and Kaplan (2018).

Yet, the MMP dataset does provide raw data pertaining to the travel dates
and sojourn times for a sizeable number of former undocumented immigrants in
the United States. This raises the question: is it possible, with proper
probability modeling, to infer the population characteristics of the
population of undocumented immigrants, such that when members of that
population are contacted via snapshot samples in Mexico, conditioning on
that sampling would reveal observations comparable to what is seen in the
actual data? If so, then viewed through the lens of the correcting model,
one could use the MMP data to learn about the entire undocumented immigrant
population.

Providing such a model and performing the attendant statistical analysis is
the goal of this paper. We construct a new probabilistic model of
undocumented migration flows back and forth across the southern border that
allows estimation of the number of migrants on either side. The model allows
for both solitary migrants, that is, those who take only a single trip to
the US over all time, and circular migrants, who make multiple repeat visits
to the US (e.g. as seasonal laborers). Given this model, we can determine
the conditional characteristics of the migrant population that would be
found in a snapshot sample administered in Mexico, and work backwards via
maximum likelihood estimation to recover the features of the entire migrant
population. While our model was developed specifically with the MMP data in
mind, our methodological approach could be applied more generally to
population snapshot samples in other settings, thus the methods are of
independent interest. Nonetheless, our numerical results based on the MMP
are of special relevance given the ongoing immigration debate in the US. To
preview our major findings, we find that from 1980 through roughly 2005,
undocumented immigration to the US across the southern border was dominated
by circular migrants who spent relatively short spells in the US. After
2005, migration came to be dominated by solitary migrants with much longer
sojourns in the US. We note that 2005 coincides with the tightening of
security along the southern border, making it more difficult to cross
undetected, as has been discussed elsewhere (Fazel-Zarandi, Feinstein and
Kaplan, 2018; Massey, Durand and Pren, 2016). Regarding the number of
undocumented immigrants in the US, the MMP data suggest that by the end of
2016, there were about 14.6 million undocumented southern border crossers
with a 95\% confidence interval running from 9.4 million and 19.8 million
persons. For several reasons detailed in Section 12, we believe this to be a
lower bound on the total number of undocumented immigrants in the US. Our
new results are thus broadly consistent with the figures reported earlier by
Fazel-Zarandi, Feinstein and Kaplan (2018), further calling into question
the more commonly reported results based on the residual method.

Our paper proceeds as follows: in the next section, we provide more
background information on the MMP dataset we will study. Then in Section 3,
we provide a high level overview of the model to follow. \ The mathematical
formulation and derivations involved in our model unfold over Sections 4
through 8. Section 9 derives in detail the likelihood functions that
correspond to the trip departure and sojourn time data involved in the MMP
surveys. Section 10 develops a novel submodel for the average number of
trips per migrant observed in an MMP sample, based on the true average
number of trips to date among circular migrants in the population as well as
the population split between solitary and circular migrants, and uses this
model to recover the level of repeat trips in the population from the number
of trips reported in the snapshot sample. Section 11 reports the complete
statistical estimation and parameterization details employed in our
analysis, while Section 12 presents our findings both graphically and
numerically in light of earlier research regarding undocumented immigration
to the United States.

\section{Data Source: Mexican Migration Project}

The Mexican Migration Project (MMP) is a rich dataset that has been used in
a variety of academic studies regarding migration to the United States from
Mexico (Durand and Massey, 2006; Massey, Durand and Pren, 2016). The MMP
began in 1982, and has administered an annual survey in Mexico since 1987.
Subjects are determined by first selecting communities based on
anthropological criteria and then sampling within those communities. On
average, four communities that are intended to be representative of four
different levels of urbanization are surveyed each year. In smaller
communities, investigators complete a full census and 200 households are
selected randomly. In more urban communities, a specific neighborhood is
chosen, from which 200 households are surveyed at random. Initially, those
communities sampled were selected from Western Mexico, where U.S. migration
is most common, but recent surveys have covered a more diverse geographic
area. The survey is conducted in the winter, since seasonal migrants are
most likely to have returned to Mexico at that time of year. Through June
2017, the MMP has collected detailed immigration histories---as well as
demographic and economic information---from 161 communities and 26,056
households in Mexico. The MMP data is publicly available online
(https://mmp.opr.princeton.edu/), and also contains information on a
smaller, non-random sample of migrants surveyed in the United States. \ In
our analysis, we focus on MMP data regarding 3,480 heads of households who
have spent time in the United States as undocumented immigrants between 1980
and 2016. \ Our analysis of these data proceeds in accord with a newly
derived migration model, which is summarized next.

\section{Model Overview}

Our model considers the relative composition, travel patterns, and sojourns
in the United States of undocumented immigrants who cross the southern
border from Mexico. As illustrated in Figure 1, there are two classes of
migrants in our model: solitary and circular. Solitary migrants travel only
once to the United States, and remain there for a random sojourn time taken
from a solitary-migrant-specific probability distribution before returning
to Mexico. Circular migrants make an initial (rookie) trip to the United
States, and sojourn for a random time taken from a circular-migrant-specific
probability distribution before returning to Mexico. Following any trip to
the US, with constant probability, circular migrants refrain from further
migration episodes, in which case they are said to be retired, or with
complementary probability, remain active migrants who will make at least one
repeat visit to the US. In our analysis, we assume the migration process
began in year $t=1$ (which corresponds to calendar year 1980), and continues
until year $\tau $; in our analysis, $\tau =37$ (which corresponds to
calendar year 2016). Time is discrete in this model. We further adopt the
following travel convention: migrant departures from Mexico to the US occur
at the start of a year, while returns from the US to Mexico occur at the end
of the year. Annual population quantities are estimated at the end of each
year, with one year sojourn times corresponding to departure from and return
to Mexico at the start and end of the same year. We will now develop the
mathematics underlying our model over the next several sections of the paper.

\begin{figure}[t]
	\includegraphics[width=0.98\columnwidth]{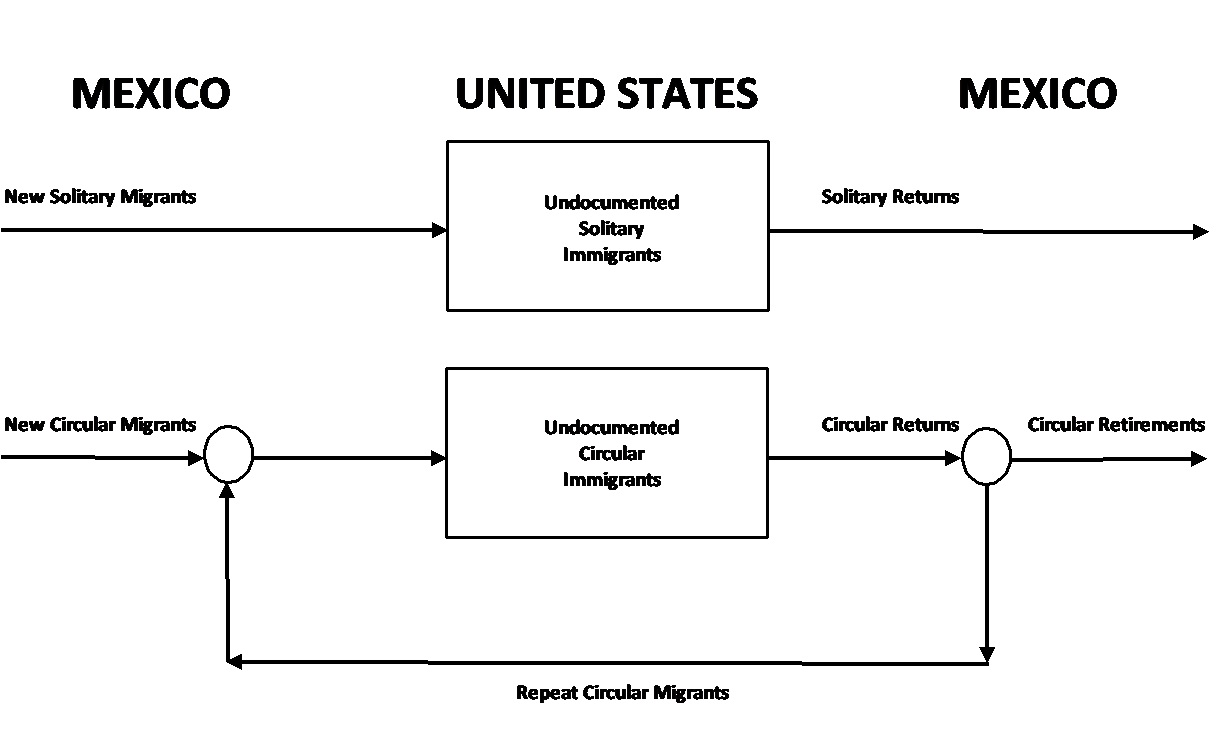}
	\caption{Model Overview}
\end{figure}

\section{Solitary Migrant Departures}

We first consider solitary migrants, that is, those who make only a single
undocumented visit to the United States over all time. From the population
of \textit{all} migrants who traveled to the United States and were
undocumented at some point between 1980 and 2016, we define $N_{S}(t)$ as
the cumulative fraction that migrated to the US between years 1980 and $%
1980+t-1$ \textit{and }consists solely of solitary migrants. \ Our timing
convention thus sets year 1980 as $t=1$, and year 2016 as $t=37\equiv \tau $%
. The total fraction of the migrant population that traveled to the United
States between 1980 and 2016 and is solitary, $N_{S}(\tau )$, is denoted by $%
\phi $, which is a parameter to be estimated from the data.

Solitary migrant departure rates are defined via the solitary migrant 
\textit{trip hazard} $h_{S}(t)$, which is defined via%
\begin{equation}
N_{S}(t)h_{S}(t)=N_{S}(t)-N_{S}(t-1)\text{, }t=1,2,...,\tau .  \label{eq2}
\end{equation}%
Note that as all migration in presumed to begin at $t=1$, we have $%
N_{S}(0)\equiv 0$ and consequently $h_{S}(1)\equiv 1$. Equation (\ref{eq2})
implies the recursion%
\begin{equation}
N_{S}(t-1)=(1-h_{S}(t))N_{S}(t)\text{, }t=1,2,...,\tau  \label{eq3}
\end{equation}%
with solution%
\begin{equation}
N_{S}(t)=N_{S}(\tau )\tprod\nolimits_{j=t+1}^{\tau }(1-h_{S}(j))\text{, }%
t=1,2,...,\tau .  \label{eq4}
\end{equation}

We also define $f_{S}(t)$, the fraction of all solitary migration trips to
the US that depart from Mexico in year $t$, as%
\begin{equation}
f_{S}(t)=\frac{N_{S}(t)h_{S}(t)}{N_{S}(\tau )}=h_{S}(t)\tprod%
\nolimits_{j=t+1}^{\tau }(1-h_{S}(j))\text{, }t=1,2,...,\tau .  \label{eq5}
\end{equation}%
This enables an alternative representation of $h_{S}(t)$ as%
\begin{equation}
h_{S}(t)=f_{S}(t)/\sum_{j=1}^{t}f_{S}(j)\text{, }t=1,2,...,\tau .
\label{eq6}
\end{equation}%
We estimate $f_{S}(t)$ from the data, yielding estimates for $h_{S}(t)$ as
well via equation (\ref{eq6}).

\section{Circular Migrant Departures}

Next we consider circular migrants. Following any stay in the United States,
circular migrants return to Mexico where, with probability $q$, they \textit{%
retire} and forever refrain from further undocumented migration, but with
probability $1-q$ remain \textit{active} and available for additional
migration visits. The parameter $q$ is estimated from the data. From the
population of \textit{all }migrants traveling between 1980 and 2016, we
define $N_{C}(t)$ as the cumulative fraction that migrated \textit{at least
once} up to and including year $t$ (including both retired and active
migrants) \textit{and }consists solely of circular migrants. The total
fraction of the migrant population that traveled to the United States
between 1980 and 2016 and is circular, $N_{C}(\tau )$, equals $1-\phi $ (as
all migrants are either solitary or circular in this model).

Unlike solitary migrants, circular migrants average more than one visit to
the US (indeed via the retirement assumption, the average number of
undocumented migration visits to the US for circular migrants is equal to $%
1/q$). Circular trip timing is heavily dependent upon the average number of
migration visits taken to date among circular migrants. We define $m_{C}(t)$
as the average number of trips taken from years one through $t$ over \textit{%
all} circular migrants who have traveled by year $t$, whether active or
retired. As we assume the migration process begins in year 1, we must have $%
m_{C}(1)=1$; also we must have $m_{C}(t)\geq 1$ as all circular migrants in
the population as of time $t$ have taken at least one trip. The parameters $%
m_{C}(t)$ will be estimated from the data.

Unlike solitary migrants, where the number of migrants equals the number of
trips, with circular migrants one must distinguish people from migration
episodes. Circular migrant departure rates are defined via the circular
migrant trip hazard $h_{C}(t)$, which is obtained via%
\begin{equation}
N_{C}(t)h_{C}(t)=N_{C}(t)m_{C}(t)-N_{C}(t-1)m_{C}(t-1)\text{, }%
t=1,2,...,\tau .  \label{eq7}
\end{equation}%
Note that the right-hand side of equation (\ref{eq7}) reports the (relative)
number of circular trip departures that occur in year $t$. Equation (\ref%
{eq7}) implies the recursion%
\begin{equation}
N_{C}(t-1)=\frac{m_{C}(t)-h_{C}(t)}{m_{C}(t-1)}N_{C}(t)\text{, }%
t=1,2,...,\tau  \label{eq8}
\end{equation}%
with solution%
\begin{equation}
N_{C}(t)=N_{C}(\tau )\frac{m_{C}(\tau )}{m_{C}(t)}\tprod\nolimits_{j=t+1}^{%
\tau }(1-\frac{h_{C}(j)}{m_{C}(j)})\text{, }t=1,2,...,\tau .  \label{eq9}
\end{equation}

We also define $f_{C}(t)$, the fraction of all circular migrant trips to the
US that depart from Mexico in year $t$, as%
\begin{eqnarray}
f_{C}(t) &=&\frac{N_{C}(t)h_{C}(t)}{N_{C}(\tau )m_{C}(\tau )}  \label{eq10}
\\
&=&\frac{h_{C}(t)}{m_{C}(t)}\tprod\nolimits_{j=t+1}^{\tau }(1-\frac{h_{C}(j)%
}{m_{C}(j)})\text{, }t=1,2,...,\tau .  \notag
\end{eqnarray}%
This enables an alternative representation of $h_{C}(t)$ as%
\begin{equation}
h_{C}(t)=\frac{m_{C}(t)f_{C}(t)}{\sum_{j=1}^{t}f_{C}(j)}\text{, }%
t=1,2,...,\tau .  \label{eq11}
\end{equation}%
We estimate $f_{C}(t)$ from the data, yielding estimates for $h_{C}(t)$ as
well via equation (\ref{eq11}).

\subsection{Active Circular Migrants: Rookies and Repeaters}

Within the circular migrant population, we have already distinguished
between active and retired migrants. Among active circular migrants, we need
to further distinguish between those embarking upon their first trip (%
\textit{rookies}), and those taking a subsequent trip (\textit{repeaters}).
At time $t$, the fraction of \textit{all }migrants who are circular rookies
by virtue of embarking on their first trip to the US, $n_{C}(t)$, is given by%
\begin{equation}
n_{C}(t)=N_{C}(t)-N_{C}(t-1)\text{, }t=1,2,...,\tau .  \label{eq12}
\end{equation}%
Recalling from equation (\ref{eq7}) the (relative) total number of circular
migrant trip departures, the number of \textit{repeat }circular departures
in year $t$, $R(t)$, is found by subtracting out the rookie trips, that is%
\begin{eqnarray}
R(t) &=&N_{C}(t)h_{C}(t)-n_{C}(t)  \label{eq13} \\
&=&N_{C}(t)(m_{C}(t)-1)-N_{C}(t-1)(m_{C}(t-1)-1)\text{, }t=1,2,...,\tau . 
\notag
\end{eqnarray}%
We assume that circular repeaters who depart for the US at the beginning of
year $t$ are selected at random from among all active circular migrants in
Mexico at the end of year $t-1$, as will be formalized shortly.

\section{Sojourn Time Distributions}

The duration of time migrants spend in the United States on any visit is
dictated by the \textit{sojourn times }distributions. The travel time
convention is that departures from Mexico to the US occur at the start of a
year, while return travel from the US to Mexico occurs at the end of a year.
Migrants who depart and return in the same year $t$ thus have a sojourn time
of one year, while in general migrants who depart in year $t$ and return in
year $t+j-1$ have sojourn times of $j$ years. Let $A_{S}$ and $A_{C}$ denote
the sojourn time for a randomly selected visit to the US by a solitary or
circular migrant respectively. The sojourn time distributions are denoted by%
\begin{equation}
a_{S}(j)=\Pr \{A_{S}=j\}\text{, }j=1,2,...37  \label{eq14}
\end{equation}%
and%
\begin{equation}
a_{C}(j)=\Pr \{A_{C}=j\}\text{, }j=1,2,...37.  \label{eq15}
\end{equation}%
Both of these distributions will be estimated from the data. Note that we
restrict sojourn times to at most equal 37 years, which is the maximum time
observable in the data (from 1980 through 2016). This is a conservative
assumption in that restricting sojourn times to fall less than or equal to
37 years will produce a lower bound on the estimated number of undocumented
immigrants in the United States.

\section{Five Migrant Subpopulations}

Of all migrants who ever traveled to the United States between 1980 and
2016, the fraction that departed up to and including year $t$ is equal to $%
N_{S}(t)+N_{C}(t)$, the sum of the (relative) number of solitary and
circular migrants who have traveled by year $t$. This fraction can be
decomposed into five mutually exclusive subpopulations:

\qquad (1) solitary migrants in the US at the end of year $t$, denoted by $%
P_{S}(t)$;

\qquad (2) solitary migrants who have returned to Mexico by the end of year $%
t$, given by $N_{S}(t)-P_{S}(t)$;

\qquad (3) circular migrants in the US at the end of year $t$, denoted by $%
P_{C}(t)$;

\qquad (4) circular migrants who have returned to Mexico and retired (i.e.
quit) by the end of year $t$, denoted by $Q(t)$; and

\qquad (5) circular migrants who have returned to Mexico and remain active
and available for further travel, given by $N_{C}(t)-P_{C}(t)-Q(t)$.

\noindent $N_{S}(t)$ and $N_{C}(t)$ have been deduced as equations (\ref{eq4}%
) and (\ref{eq9}) respectively. It remains to specify $P_{S}(t)$, $P_{C}(t)$%
, and $Q(t)$.

\subsection{Solitary Migrants in the United States}

The fraction of all migrants who are solitary and residing in the United
States at the end of year $t$ follows from the convolution of solitary
departures with the solitary sojourn time distribution, and is given by%
\begin{equation}
P_{S}(t)=\sum_{j=1}^{t}N_{S}(j)h_{S}(j)\Pr \{A_{S}>t-j+1\}\text{, }%
t=1,2,...,\tau   \label{eq16}
\end{equation}%
where%
\begin{equation}
\Pr \{A_{S}>t\}=\sum_{j=t+1}^{37}a_{S}(j)\text{, }t=1,2,...,37.  \label{eq17}
\end{equation}%
Note that equation (\ref{eq16}) reflects our travel time convention, one
implication of which is that to be resident in the United States at the end
of any year requires a sojourn time strictly greater than one year.

\subsection{Circular Migrants in the United States}

The circular migrant population residing in the United States at the end of
year $t$ is similarly defined as%
\begin{equation}
P_{C}(t)=\sum_{j=1}^{t}N_{C}(j)h_{C}(j)\Pr \{A_{C}>t-j+1\}\text{, }%
t=1,2,...,\tau .  \label{eq18}
\end{equation}%
Note that equation (\ref{eq18}) accounts for both rookie and repeat circular
migrants as discussed earlier.

\subsection{Retired Circular Migrants in Mexico}

The relative size of the retired circular migrant population can be easily
computed by noting that since the total number of circular migrant
departures from Mexico up to and including year $t$ equals $N_{C}(t)m_{C}(t)$%
, the relative number of \textit{completed }circular migrant visits to the
United States must equal $N_{C}(t)m_{C}(t)-P_{C}(t)$, since those circular
migrants in the US at the end of year $t$ have obviously not yet completed
their most recent trip. Given that any circular migrant retires with
probability $q$ following return to Mexico, the fraction of all migrants who
are circular and have retired from further migration by the end of year $t$, 
$Q(t)$, is given by

\begin{equation}
Q(t)=(N_{C}(t)m_{C}(t)-P_{C}(t))q\text{, }t=1,2,...,\tau .  \label{eq19}
\end{equation}

\section{Active Circular Migrant Behavior}

Next we detail how we model the repeat-trip taking behavior of active
circular migrants. Equation (\ref{eq13}) presents the frequency of repeat
trips at the start of year $t$, $R(t)$, as the difference between total and
rookie circular migrant departures. We assume that the active circular
migrants who embark on such repeat trips at the start of year $t$ are
selected at random from active circular migrants resident in Mexico at the
end of year $t-1$. More formally, let $\rho (t)$ be the repeat trip
probability at the start of year $t$. We define $\rho (t)$ as%
\begin{equation}
\rho (t)=\frac{R(t)}{N_{C}(t-1)-P_{C}(t-1)-Q(t-1)}\text{, }t=1,2,...,\tau .
\label{eq20}
\end{equation}%
$\rho (t)$ is thus the conditional probability that an \textit{active}
circular migrant at the end of year $t-1$ embarks on a repeat migration
visit to the United States in year $t$. An important consequence is that
having returned to Mexico at the end of, say, year $t-j$, the probability an
active circular migrant \textit{refrains} from subsequent travel from year $%
t-j+1$ to year $t$, $\nu (t-j+1,t)$, is given by%
\begin{equation}
\nu (t-j+1,t)=\tprod\nolimits_{k=t-j+1}^{t}(1-\rho (k))\text{, }j=1,2,...,t%
\text{; }t=2,3,...,\tau .  \label{eq21}
\end{equation}%
Since the migration process begins in year $1$, no repeat travel is possible
in year 1, that is, $\rho (1)=0$. Equation (\ref{eq21}) will prove to be
very important in writing down the likelihood of observing trip departure
and return data for circular migrants.

\section{Linking Migration Model to Data: Trip Hazard/Sojourn Time Data
Likelihood}

As stated in the introduction, the Mexican Migration Project data used to
calibrate our model derive from 3,480 heads of households who have spent
time in the United States as undocumented immigrants between 1980 and 2016.
The MMP \textquotedblleft mig161\textquotedblright\ file contains
information on up to 25 border crossing attempts for each head of household
as well as the departure year and duration of his or her \textit{most recent
trip} to the US. These data were gleaned from 30 surveys conducted from 1987
through 2016.\ We record the departure year for each head of household in
the data. Trip durations are reported in months, which we round up to the
nearest full year in defining a migrant's \textit{sojourn time}.

From these 3,480 individual observations, we derive frequency counts of the
number of migrants taking first trips (which could be attributed to either
solitary or circular migrants), or repeat trips (which could only be
attributed to circular migrants) indexed by the year of departure, sojourn
time, and survey year associated with each observation. \ Recall our time
convention of representing the years 1980 through 2016 by $t=1,...,37$. The
earliest of the 30 surveys analyzed was administered in 1987, corresponding
to $t=8$. \ Focus on a survey administered in year $t$, $t=8,9,...,37$. For
a migrant observed in this survey, let $i$ denote the departure year to the
US counted \textit{backwards }from the survey year while respecting our
travel convention governing start-of-year departures to the US and
end-of-year returns to Mexico, and $j$ denote the sojourn time spent in the
United States (again respecting our travel convention that departure at the
start of a year and return at the end of the same year yields a sojourn time 
$j=1$ year). We define $s_{ijt}$ as the number of migrants in the data
collected in year $t$ who made a single trip (since 1980) that departed in
year $t-i+1$ and sojourned for $j$ years in the US. We similarly define $%
r_{ijt}$ as the number of migrants whose most recent repeat trip departed in
year $t-i+1$, sojourned for $j$ years in the US, and was recorded in the
year $t$ MMP survey. We are able to determine whether a trip is a repeat
visit or not from the total number of migration episodes recorded for each
individual in the data; if this number equals unity, the trip observed is
the first migration episode, while if more than one trip has been recorded,
we know that the most recent visit must be a repeat trip.

\noindent \qquad Presume that the true average numbers of trips per circular
migrant from years 1 though $t$, $m_{C}(t)$, are fixed and known. Also
presume knowledge of the trip hazards and sojourn time probabilities for
both solitary and circular migrants, the fraction of all migrants in the
population who are solitary ($\phi =N_{S}(\tau )$), and the retirement
probability following any circular migrant trip to the US ($q$). A migrant
can only be included in an MMP survey administered in year $t$ if that
migrant was \textit{in Mexico} that year. The fraction of all migrants who
were in Mexico in year $t$ and hence available for MMP sampling, $\sigma (t)$%
, can be computed directly via equations (\ref{eq4}), (\ref{eq9}), (\ref%
{eq16}) and (\ref{eq18}) as%
\begin{equation}
\sigma (t)=N_{S}(t)-P_{S}(t)+N_{C}(t)-P_{C}(t)\text{, }t=8,9,...,37.
\label{eq22}
\end{equation}%
Conditional upon being in Mexico in year $t$, the probability that an MMP
sampled migrant was \textit{solitary}, departed to the US at the start of
year $t-i+1$, and sojourned for $j\leq i$ years before returning to Mexico
at the end of year $t-i+j$, $\mathcal{L}_{S}(i,j,t)$, is given by%
\begin{equation}
\mathcal{L}_{S}(i,j,t)=\frac{\phi f_{S}(t-i+1)a_{S}(j)}{\sigma (t)}\text{, }%
\begin{array}{c}
t=8,9,...,37;\text{ } \\ 
i=1,2,...,t; \\ 
j=1,2,...,i.%
\end{array}
\label{eq23}
\end{equation}%
\qquad \qquad

The conditional probability that an MMP sampled migrant in year $t$ was a 
\textit{circular repeater}, most recently departed to the US at the start of
year $t-i+1$, sojourned for $j\leq i$ years before returning to Mexico at
the end of year $t-i+j$, and remained in Mexico from the start of year $%
t-i+j+1$ to year $t$, $\mathcal{L}_{C}^{>1}(i,j,t)$, is given by%
\begin{eqnarray}
\mathcal{L}_{C}^{>1}(i,j,t)=\frac{R(t-i+1)a_{C}(j)\{q+(1-q)\nu (t-i+j+1,t)\}%
}{\sigma (t)} &&\text{, }  \notag \\
\begin{array}{c}
t=8,9,...,37;\text{ } \\ 
i=1,2,...,t-1; \\ 
j=1,2,...,i.%
\end{array}
&&  \label{eq24.1}
\end{eqnarray}%
Note that the index $i$ runs to $t-1$ since if the migration process began
in 1980, it is not possible for a trip in that year to be a repeat trip
(equivalently $\mathcal{L}_{C}^{>1}(t,j,t)=0$) . Also note that for the
observed departure and sojourn times to correspond to the circular migrant's 
\textit{most recent trip}, it must be that the migrant in question did not
travel between years $t-i+j+1$ and $t$ inclusive, for had such travel
occurred, then the observed trip could not have been that migrant's most
recent! Finally, note that there are two ways a returned circular migrant
could remain in Mexico until year $t$ post-return in year $t-i+j$: either by
retiring from further travel (with probability $q$), or by remaining active
(with probability $1-q$) but refraining from repeat travel between years $%
t-i+j+1$ and $t$ (with probability $\nu (t-i+j+1,t)$).

Finally, the conditional probability that an MMP sampled migrant in year $t$
was a \textit{circular rookie}, traveled to the US at the start of year $%
t-i+1$, sojourned for $j\leq i$ years before returning to Mexico at the end
of year $t-i+j$, and remained in Mexico from the start of year $t-i+j+1$ to
year $t$, $\mathcal{L}_{C}^{1}(i,j,t)$, is given by%
\begin{eqnarray}
\mathcal{L}_{C}^{1}(i,j,t)=\frac{n_{C}(t-i+1)a_{C}(j)\{q+(1-q)\nu
(t-i+j+1,t)\}}{\sigma (t)} &&\text{,}\text{ }  \notag \\
\begin{array}{c}
t=8,9,...,37;\text{ } \\ 
i=1,2,...,t; \\ 
j=1,2,...,i.%
\end{array}
&&  \label{eq25}
\end{eqnarray}

Via the migration population model, 
\begin{equation}
\sum_{i=1}^{t}\sum_{j=1}^{i}(\mathcal{L}_{S}(i,j,t)+\mathcal{L}%
_{C}^{>1}(i,j,t)+\mathcal{L}_{C}^{1}(i,j,t))=1\text{, }t=8,9,...,37,
\label{eq26}
\end{equation}%
for every migrant sampled is either solitary, a circular repeater, or a
circular rookie.

To build the data likelihood of observing the 3,480 departure and sojourn
times sampled, note that the conditional probability of observing a
confirmed circular repeater is given by $\mathcal{L}_{C}^{>1}(i,j,t)$, while
the conditional probability of observing a migrant who made a single trip
equals $\mathcal{L}_{S}(i,j,t)+\mathcal{L}_{C}^{1}(i,j,t)$, the sum of the
solitary and circular rookie probabilities. \ Presuming independence across
the MMP samples, the resulting \textit{trip departure/sojourn time} data
likelihood, $\mathcal{L}_{\mathbf{f,a}}$, is thus given by%
\begin{equation}
\mathcal{L}_{\mathbf{f,a}}=\tprod\nolimits_{t=8}^{37}\tprod%
\nolimits_{i=1}^{t}\tprod\nolimits_{j=1}^{i}(\mathcal{L}_{S}(i,j,t)+\mathcal{%
L}_{C}^{1}(i,j,t))^{s_{ijt}}\mathcal{L}_{C}^{>1}(i,j,t)^{r_{ijt}}.
\label{eq27}
\end{equation}

\section{Linking Migration Model to Data: Mean Trip Calibration}

The migration model derived requires estimation of $m_{C}(t)$, the mean
number of trips taken among all circular migrants who have traveled at any
time by the end of year $t$. The MMP data include the number of trips since
1980 for each migrant in the sample. Estimating $m_{C}(t)$ directly from
these individual level data is not straightforward, however, due to the fact
that all survey respondents were in Mexico at the time of sampling, and in
addition one can not distinguish between solitary migrants or circular
migrants who have only taken a single trip. Rather than trying to recover
the probability distribution of the number of trips taken by sampled
individuals, we instead appeal to the central limit theorem to model the
probability distribution of the conditional \textit{average} number of trips
taken to date among all migrants by time $t$, given that this average is
taken over sampled migrants, whether solitary or circular, who are in Mexico
as of the end of year $t$. The process by which this is accomplished is
referred to as \textit{mean trip calibration}. As will be shown, the
approximating normal distribution for the sampled average number of trips
among survey respondents in a given year depends in a complicated yet
tractable manner on the parameters $m_{C}(t)$ that we seek to estimate. This
normal distribution also serves as the data likelihood for the observed
conditional average number of trips sampled in the MMP, which when
multiplied by the trip departure/sojourn time likelihood (which itself is
conditional upon parameters $m_{C}(t)$) creates an overall likelihood
function from which maximum likelihood estimates can be computed.

The idea behind mean trip calibration is to first compute (or approximate)
the mean number of trips to date in each of the three circular migrant
subpopulations: circular migrants in the US, retired circular migrants in
Mexico, and active circular migrants in Mexico. Once these quantities have
been determined, the parameters $m_{C}(t)$ can be computed as a weighted
average over the size of these three circular subpopulations. Finally,
recognizing that solitary migrants always average one trip (whether in the
US or Mexico), it is possible to produce the average number of trips one
would find from a weighted average across the solitary, circular retired and
circular active populations in Mexico, providing the data link to the
unknown required parameters $m_{C}(t)$.

\subsection{Mean Trips Among Circular Migrants in the United States}

Recall that in year $t$, there are $N_{C}(t)h_{C}(t)=(1-\phi )m_{C}(\tau
)f_{C}(t)$ circular migrant trips to the US in total, of which $R(t)$ are
repeat trips, and the remaining $n_{C}(t)=N_{C}(t)-N_{C}(t-1)$ are rookie
trips. Let the (as yet unknown) average number of trips to date among 
\textit{active circular repeaters in Mexico} at the end of year $t$ be
denoted by $m_{C|Active}(t)$. Then the total number of trips to date taken
by all circular migrants who travel to the US in year $t$, including year $t$
trips, equals $n_{C}(t)+R(t)(1+m_{C|Active}(t-1))$. At the end of year $t$,
there are $P_{C}(t)$ circular migrants in the United States, and the total
number of trips to date experienced by those migrants must be given by the
total trips those migrants \textquotedblleft carried\textquotedblright\ with
them to the US in each of years 1 through $t$. This implies that the average
number of trips to date among circular migrants in the US at the end of year 
$t$, $m_{C|US}(t)$, must equal%
\begin{eqnarray}
m_{C|US}(t) &=&\frac{1}{P_{C}(t)}%
\sum_{i=1}^{t}(n_{C}(i)+R(i)(1+m_{C|Active}(i-1)))\Pr \{A_{C}>t-i+1\}
\label{eq28} \\
&=&1+\frac{1}{P_{C}(t)}\sum_{i=1}^{t}(R(i)m_{C|Active}(i-1))\Pr
\{A_{C}>t-i+1\}\text{, }t=1,2,...,\tau .  \notag
\end{eqnarray}%
We have thus shown how, given the average number of trips to date among
active circular repeaters in Mexico, one can compute the average number of
trips to date among circular migrants in the United States.

\subsection{Mean Trips Among Retired Circular Migrants}

Let $D_{C}(t)$ denote the relative number of circular migrant returns from
the US back to Mexico at the end of year $t$. A fraction $q$ of these
returns retire, thus the population of retired migrants in Mexico at the end
of year $t$, $Q(t)$, evolves according to%
\begin{equation}
Q(t)=Q(t-1)+qD_{C}(t)\text{, }t=1,2,...,\tau .  \label{eq29}
\end{equation}%
However, $Q(t)$ is already known from equation (\ref{eq19}), which means
that we can recover the number of circular returns from Mexico in year $t$ as%
\begin{equation}
D_{C}(t)=\frac{Q(t)-Q(t-1)}{q}\text{, }t=1,2,...,\tau .  \label{eq30}
\end{equation}%
In a manner similar to equation (\ref{eq28}), we calculate the average
number of trips to date among \textit{retired circular migrants}, denoted $%
m_{C|Q}(t)$, by carrying the average trips to date among returning circular
migrants from the US. While in principle we could make a more complicated
exact calculation by keeping track of circular migrant arrival and sojourn
times in the US, instead we approximate by treating returning circular
migrants as randomly sampled inside the US. We thus employ the
return-weighted average of trips to date among circular migrants in the US
to approximate $m_{C|Q}(t)$, that is, we set%
\begin{equation}
m_{C|Q}(t)=\frac{\sum_{i=1}^{t}m_{C|US}(i)D_{C}(i)q}{Q(t)}\text{, }%
t=1,2,...,\tau .  \label{eq31}
\end{equation}%
We expect this approximation to work well for as will be seen, circular
migrant sojourn times tend to be short. Equation (\ref{eq31}) thus shows how
we compute the average number of trips to date among retired circular
migrants from mean trips to date among circular migrants in the US.

\subsection{Mean Trips Among Active Circular Migrants in Mexico}

To complete mean trip calibration requires computing the mean number of
trips to date among active circular migrants in Mexico, $m_{C|Active}(t)$.
Suppose that we know the average number of trips to date, $m_{C}(t)$, over 
\textit{all} $N_{C}(t)$ circular migrants. Clearly $m_{C}(t)$ must be the
population-weighted average of the conditional average number of trips to
date over the three circular migrant subpopulations, that is, 
\begin{eqnarray}
m_{C}(t)=\frac{%
m_{C|US}(t)P_{C}(t)+m_{C|Q}(t)Q(t)+m_{C|Active}(t)(N_{C}(t)-P_{C}(t)-Q(t))}{%
N_{C}(t)} &&\text{, }  \notag \\
t=1,2,...,\tau . &&  \label{eq32}
\end{eqnarray}

\subsection{Solving\ The Mean Trip Calibration Equations}

Given $m_{C}(t)$, equations (\ref{eq28}), (\ref{eq31}) and (\ref{eq32}) can
be iteratively resolved to identify the three subpopulation averages $%
m_{C|US}(t)$, $m_{C|Q}(t)$, and $m_{C|Active}(t)$. Note that in 1980 which
corresponds to $t=1$, the number of circular repeat migrant trips $R(1)=0$,
which via equation (\ref{eq28}) forces $m_{C|US}(1)=1$. This in turn forces $%
m_{C|Q}(1)=1$ via equation (\ref{eq31}), and since $m_{C}(1)=1$ by
definition, equation (\ref{eq32}) forces $m_{C|Active}(1)=1$ as well. From $%
t=2$ onwards, one simply cycles through equations (\ref{eq28}), (\ref{eq31})
and (\ref{eq32}) in order to determine $m_{C|US}(t)$, $m_{C|Q}(t)$, and $%
m_{C|Active}(t)$ for all $t=2,3,...,37$.

\subsection{Mean Trips Among Migrants In Mexico}

Recall that the MMP data enable estimation of the mean number of trips to
date among migrants sampled in Mexico, which includes active and retired
circular migrants as well as returned solitary migrants. The population
average number of trips to date over migrants in Mexico at the end of year $%
t $, $m_{Mexico}(t)$, is again a population-weighted average, but now over
the migrant populations in Mexico. As trips to date among solitary migrants
always equal one, we have%
\begin{eqnarray}
m_{Mexico}(t)=\frac{1\times
(N_{S}(t)-P_{S}(t))+m_{C|Q}(t)Q(t)+m_{C|Active}(t)(N_{C}(t)-P_{C}(t)-Q(t))}{%
N_{S}(t)-P_{S}(t)+N_{C}(t)-P_{C}(t)}\text{,} &&  \notag \\
t=1,2,...,\tau . &&  \label{eq33}
\end{eqnarray}%
Equation (\ref{eq33}) thus provides the mean trips to date among migrants in
Mexico, which is also the \textit{expected} average trips to date in a
random sample of migrants in Mexico, providing the link from the migration
model to the number of trips to date reported by respondents in the MMP as
described next.

\subsection{Data Likelihood For Mean Trip Calibration}

Via appeal to the central limit theorem, we assume that the observed sample
average of trips to date across survey respondents in a year $t$ MMP survey, 
$\overline{m}_{Mexico}(t)$, is drawn from a normal distribution with mean $%
m_{Mexico}(t)$. Rather than attempt to derive a population-based estimator
for the variance of such sample means, we will simply use the squared
standard error of the observed sample mean, $s_{Mexico}^{2}(t)$, as our
estimate for the variance of the sampling distribution (note that this
reflects the different sample sizes in different MMP surveys). Presuming
independence across the MMP samples as before, the \textit{mean trip
calibration }data likelihood $\mathcal{L}_{\mathbf{m}}$ is given by%
\begin{equation}
\mathcal{L}_{\mathbf{m}}=\tprod\nolimits_{t=8}^{37}\mathcal{N}(\overline{m}%
_{Mexico}(t)|m_{Mexico}(t),s_{Mexico}^{2}(t))  \label{eq34}
\end{equation}%
where $\mathcal{N}(x|\mu ,\sigma ^{2})$ is the value of a normal density
with mean $\mu $ and variance $\sigma ^{2}$ evaluated at $x$.

\section{Maximum Likelihood Estimation and Parameterization}

The total likelihood function to be maximized is

\begin{equation}
\mathcal{L=\mathcal{L}}_{\mathbf{f,a}}\mathcal{L}_{\mathbf{m}}.  \label{eq35}
\end{equation}%
To accomplish this, as is common in practice, we maximized the logarithm of
the likelihood function, given by%
\begin{eqnarray}
\log \mathcal{L} &=&\sum_{t=8}^{37}\sum_{i=1}^{t}\sum_{j=1}^{i}s_{ijt}\log (%
\mathcal{L}_{S}(i,j,t)+\mathcal{L}_{C}^{1}(i,j,t))+r_{ijt}\log \mathcal{L}%
_{C}^{>1}(i,j,t)  \notag \\
&&+\sum_{t=8}^{37}\log \mathcal{N}(\overline{m}%
_{Mexico}(t)|m_{Mexico}(t),s_{Mexico}^{2}(t)).  \label{eq36}
\end{eqnarray}%
This maximization is over the parameters $\phi $, $q$, $f_{S}(t)$, $a_{S}(t)$%
, $f_{C}(t)$, $a_{C}(t)$, and $m_{C}(t)$, $t=1,2,...,37$. Individual sojourn
time probabilities are estimated for $a_{S}(t)$ (respectively $a_{C}(t)$)
for $t=1,2,...,20$. Owing to data sparsity, we constrain the sojourn time
probabilities $a_{S}(t)$ (respectively $a_{C}(t)$) to equal the values $%
a_{S}^{>20}$ (respectively $a_{C}^{>20}$) for $t=21,22,...,37$. All
probabilities are constrained to be non-negative, with%
\begin{equation}
\sum_{t=1}^{37}f_{S}(t)=\sum_{t=1}^{37}a_{S}(t)=\sum_{t=1}^{37}f_{C}(t)=%
\sum_{t=1}^{37}a_{C}(t)=1,  \label{eq37}
\end{equation}%
and $0\leq \phi $, $q\leq 1$. To smooth random fluctuations resulting from
sampling the number of trips to date in the MMP, we model $m_{C}(t)$ using
four piecewise-linear functions with knots at $t=$ 1, 10, 19, 28 and 37,
corresponding to calendar years 1980, 1989, 1998, 2007, and 2016. Given our
assumption that the migration process began in 1980, we have $m_{C}(1)=1$ by
definition. The number of free parameters estimated by type thus equals 40
(20 for each of the two sojourn time distributions) + 72 (36 for each of the
two trip departure distributions) + 4 (for the knots in $m_{C}(t)$) + 2 ($%
\phi $ and $q$) for a total of 118.

Maximizing (\ref{eq36}) was accomplished numerically via the Analytic Solver
Platform's Large-Scale GRG Solver Engine (https://www.solver.com/category/

\noindent product/solver-sdk-platform). As a check on the solution obtained,
we also formulated and maximized the equivalent \textquotedblleft Poisson
trick\textquotedblright\ likelihood for this problem (Baker, 1994), and
obtained identical results. As dictated by the asymptotics of maximum
likelihood estimation, the covariance matrix of the model parameters was
estimated by the negative inverse of the Hessian matrix of the maximized
log-likelihood function (Cox and Hinkley, 1974), while estimated variances
for functions of estimated parameters such as the undocumented populations
or mean sojourn times for solitary and circular migrants were obtained via
the delta method (Bishop, Fienberg and Holland, 1975).

\section{Results}

The estimated fraction of all migrants who are solitary is given by $%
\widehat{\phi }=0.63$ (95\% CI: $0.52,0.74$). However, the mean number of
circular migrant trips made by the end of 2016 is estimated as $\widehat{m}%
_{C}(37)=2.09$ (95\% CI: $1.91$, $2.27$), so while circular migrants only
account for a bit over a third of all migrants, they are estimated to be
responsible for a fraction%
\begin{equation}
\frac{(1-\widehat{\phi })\widehat{m}_{C}(37)}{\widehat{\phi }+(1-\widehat{%
\phi })\widehat{m}_{C}(37)}=\frac{0.37\times 2.09}{0.63+0.37\times 2.09}%
\approx 0.55  \label{eq38}
\end{equation}%
or 55\% of all border crossings. The post-return circular migrant retirement
probability is estimated as $\widehat{q}=0.34$ (95\% CI: $0.29,0.39$), so on
average repeat migrants complete about 3 migration episodes before retiring.
Note that this exceeds the estimated mean trips to date among circular
migrants by 2016, as it must since by the end of the study, the fraction of
circular migrants who have retired only equals $\widehat{Q}(37)/(1-\widehat{%
\phi })=0.23/0.37\approx 0.63$.

Figure 2 reports the observed ($\overline{m}_{Mexico}(t)$) and via equation (%
\ref{eq33}) the estimated ($\widehat{m}_{Mexico}(t)$) mean number of trips
to date among migrants sampled in Mexico. The piecewise linear model imposed
upon the unconditional mean circular migrant trips to date ($m_{C}(t)$)
results in model estimates that indeed smooth the sample means while
maintaining their overall time trend. Since the trip distribution likelihood 
$\mathcal{L}_{\mathbf{m}}$ employs the squared standard error of the
observed sample mean trips to date as the underlying measure of variability,
the model fits the data better in those years with smaller rather than
larger observed standard errors, which in turn imposes tighter model fits in
years with larger rather than smaller samples.

\begin{figure}[t]
	\includegraphics[width=0.98\columnwidth]{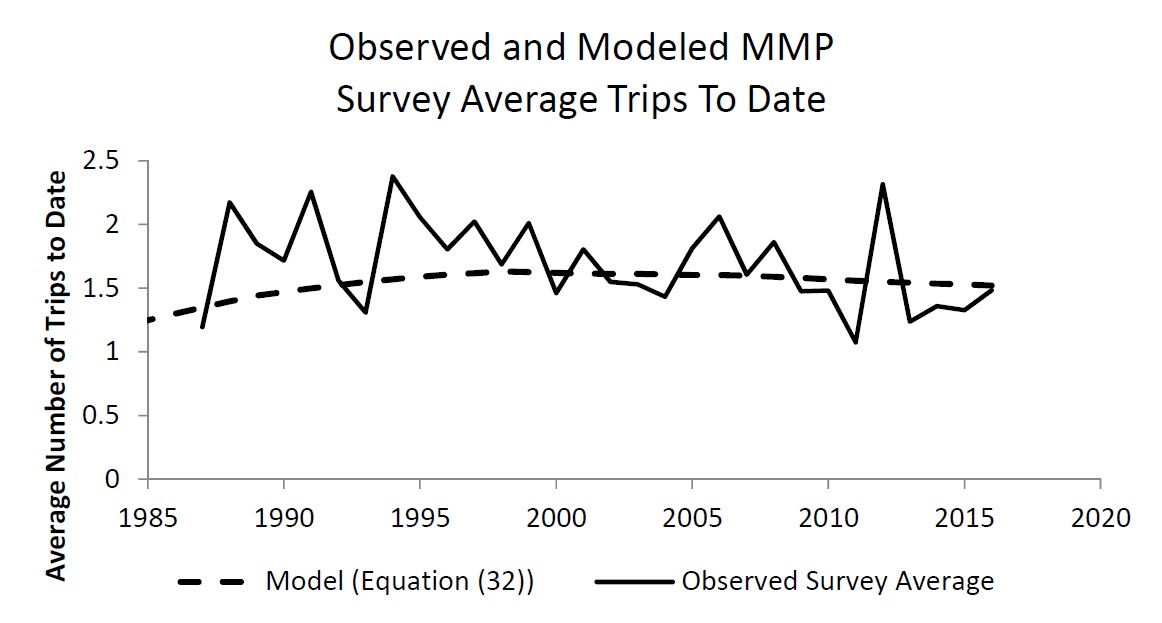}
	\caption{Mean Trip Calibration}
\end{figure}

Figure 3 presents the observed and expected in-sample trip
departure date frequency distributions (with the expected frequencies
computed by conditioning on the sample size for single and repeat travelers
in each of the 30 MMP samples, and distributing trips in accord with the
estimated trip departure time probabilities), while Figure 4 reports
observed and expected in-sample sojourn time frequency distributions (with
expected frequencies again computed based on observed sample sizes). These
figures further demonstrate the fit of the model to the data observed.

\begin{figure}[t]
	\includegraphics[width=0.98\columnwidth]{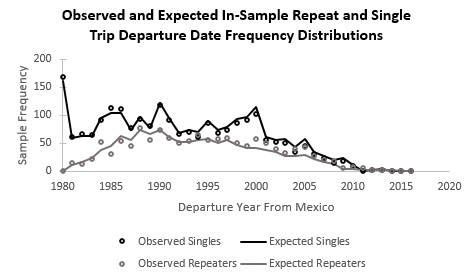}
	\caption{Observed and expected in-sample trip departure date frequency distributions}
\end{figure}

\begin{figure}[t]
	\includegraphics[width=0.98\columnwidth]{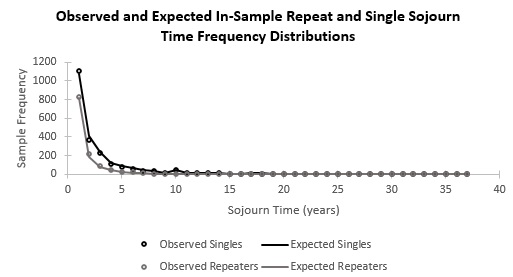}
	\caption{Observed and expected in-sample sojourn time frequency distributions}
\end{figure}

To appreciate the implications of the trip departure and sojourn
time distributions estimated by our model, we will scale the model so that
the total number of southern border crossings between 1990 and 2016 match
the 38.8 million southern border crossings estimated in Fazel-Zarandi,
Feinstein and Kaplan (2018) over this same period. That estimate, which
relied upon backcalculating border crossings from Department of Homeland
Security (DHS) apprehensions data from 1990-2004 and using published DHS
southern border crossing estimates from 2005-2016, is recognized as an
underestimate, in that the 39\% apprehension rate presumed in Fazel-Zarandi,
Feinstein and Kaplan (2018) is much higher than reported by others (e.g.
Massey, Durand and Pren, 2016), implying that fewer migrants successfully
crossed the border.

The total relative number of border crossings between 1980 and 2016 in our
model is given by the sum of solitary plus circular migrant crossings, which
equals $\phi +(1-\phi )m_{C}(37)$. Of these, the relative number of solitary
crossings that occurred between 1990 and 2016 equals $N_{S}(37)-N_{S}(10)$,
while the relative number of circular crossings during that same time period
equals $m_{C}(37)N_{C}(37)-m_{C}(10)N_{C}(10)$. Letting $T$ denote the total
number of border crossings scaled to coincide with Fazel-Zarandi, Feinstein
and Kaplan (2018), we set 
\begin{equation}
T\times \frac{\left( \widehat{N}_{S}(37)-\widehat{N}_{S}(10)\right) +\left( 
\widehat{m}_{C}(37)\widehat{N}_{C}(37)-\widehat{m}_{C}(10)\widehat{N}%
_{C}(10)\right) }{\widehat{\phi }+(1-\widehat{\phi })\widehat{m}_{C}(37)}%
=38.8\text{ million}  \label{eq39}
\end{equation}%
which yields $T=45.43$ million border crossings in total. With this scaling,
the number of solitary and circular migrant border crossings in year $t$, $%
b_{S}(t)$ and $b_{C}(t)$, are given by%
\begin{equation}
b_{S}(t)=T\frac{\phi }{\phi +(1-\phi )m_{C}(37)}f_{S}(t)\text{, }t=1,2,...,37
\label{eq40}
\end{equation}%
and%
\begin{equation}
b_{C}(t)=T\frac{(1-\phi )m_{C}(37)}{\phi +(1-\phi )m_{C}(37)}f_{C}(t)\text{, 
}t=1,2,...,37  \label{eq41}
\end{equation}%
respectively.

Figure 5 reports the resulting estimated number of southern border crossings
over time, and the figure tells a clear story. Prior to 2005, the
overwhelming majority of migration trips to the US were taken by circular
migrants. Many of these circular visits were short; our model estimates that
67\% of circular sojourn times were at most two years. Southern border
security during these years was relatively lax compared to later years. As
discussed in Massey, Durand and Pren (2016) and also in DHS reports (Roberts
et al, 2010), new fencing and technologies such as sensors and night vision
systems were introduced around 2005, while the prevalence of border patrols
also increased. In addition, the consequences of detection have increased;
prior to 2005, apprehended undocumented migrants were allowed to return
\textquotedblleft voluntarily\textquotedblright\ to Mexico with no further
punishment. Prior to 2005, the voluntary return rate to Mexico was 98\%; by
2010 this rate had fallen to 84\%\ (Fazel-Zarandi, Feinstein and Kaplan,
2018). Given this increase in border security, migrants who were able to
successfully enter the United States were more likely to remain for longer
periods of time. Indeed, our model estimates a mean sojourn time of 18.2
years (standard error 2.4 years) for solitary migrants, in contrast to the
estimated mean sojourn time of 5.8 years (standard error 0.6 years) for
circular migrant visits. Simply stated, the increase in border security has
made circular migration much more difficult, resulting in the rise in
solitary migration estimated in the data. However, further tightening of
security at the border has greatly reduced the border crossing rates for
solitary migrants as well. \ This overall decline in southern border
crossings is well-reflected in DHS data, with only 420,000 undocumented
crossings estimated for 2016 (Bailey, 2016).

\begin{figure}[t]
	\includegraphics[width=0.98\columnwidth]{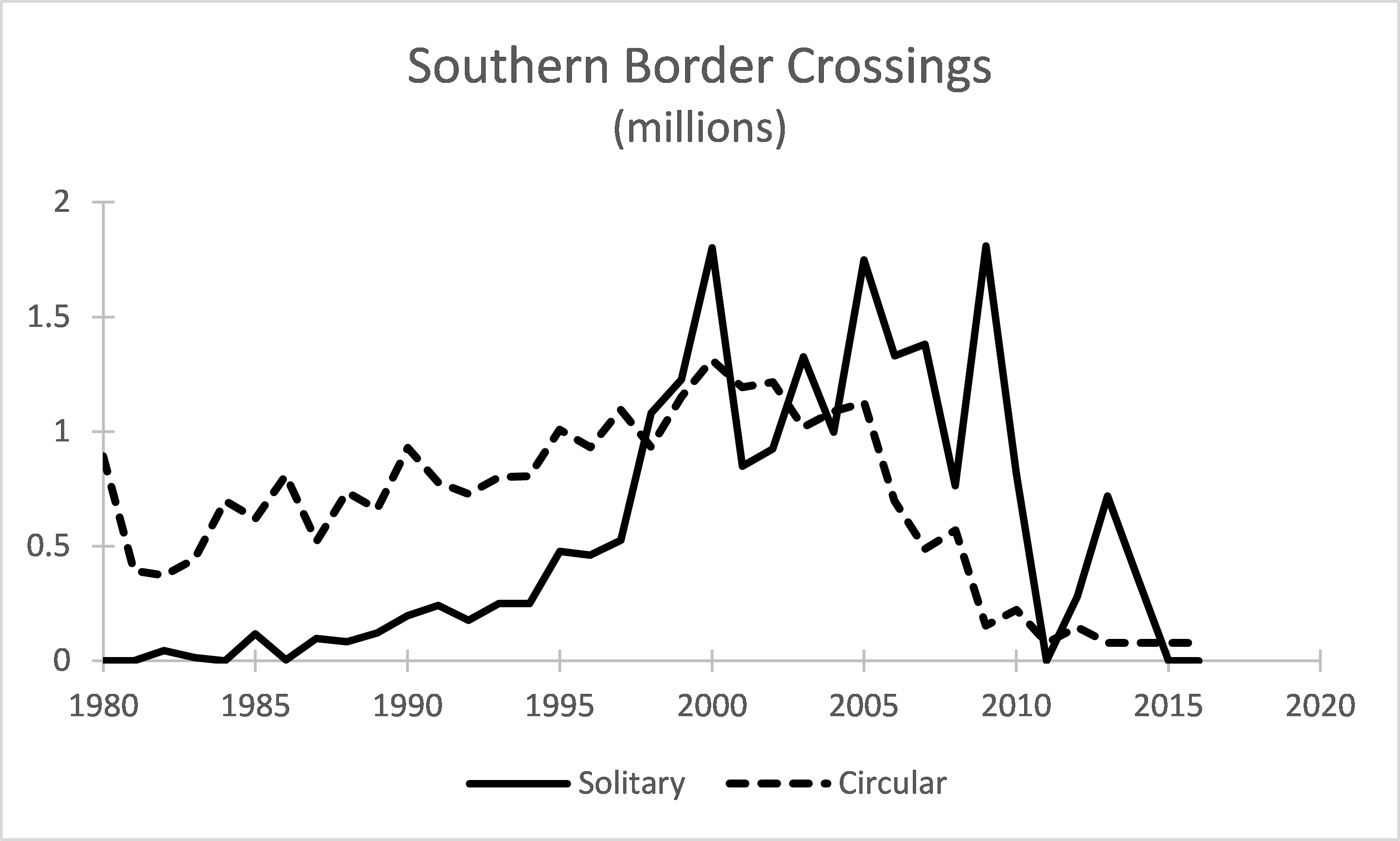}
	\caption{Solitary versus circular southern border crossings}
\end{figure}

\medskip The sojourn time undocumented migrants spend in the United States,
most commonly reflected via sojourn time-dependent emigration (hazard) rates
in the literature, has itself been a matter of some controversy. All
researchers agree that these emigration probabilities decline with the
duration of migrant stay in the United States. Based on a literature review,
Fazel-Zarandi, Feinstein and Kaplan (2018) estimated that these
probabilities fell from 40\% following one year, to 4\% after each of the
next nine years, asymptoting to 1\% annually for each year beyond ten.
Migration Policy Institute (MPI) researchers claimed that these rates were
too low, presenting rates of 50.2\%, 20.2\% and 16.6\% over these same time
periods based on their analysis of the same MMP data we have examined in
this paper (Capps \textit{et al}, 2018). Neither of these estimates
distinguish between solitary and circular migrants. Figure 6 reports the
original estimates presented by Fazel-Zarandi, Feinstein and Kaplan (2018),
the estimates proposed by Capps \textit{et al} (2018), and in addition the
uncorrected emigration rates resulting from the raw sample sojourn time
frequencies in the MMP household survey, and finally the overall emigration
hazards that result from our model for a randomly selected trip to the US
(with circular visits accounting for about 55\% of all sojourn times in
accord with equation (\ref{eq38})). The results are revealing: the
emigration rates presented by Capps \textit{et al }(2018) are very close to
what one would obtain from hazards based on the raw sojourn time frequencies
in the MMP data. However, as emphasized throughout this paper, the MMP data
constitute a snapshot sample and are thus heavily biased towards the
observation of shorter sojourn times, for survey respondents must have
returned to Mexico in time for the survey. Properly accounting for this bias
and mixing over solitary and circular visits results in emigration rates
that are much closer to those originally postulated by Fazel-Zarandi, Kaplan
and Feinstein (2018).

\begin{figure}[t]
	\includegraphics[width=0.98\columnwidth]{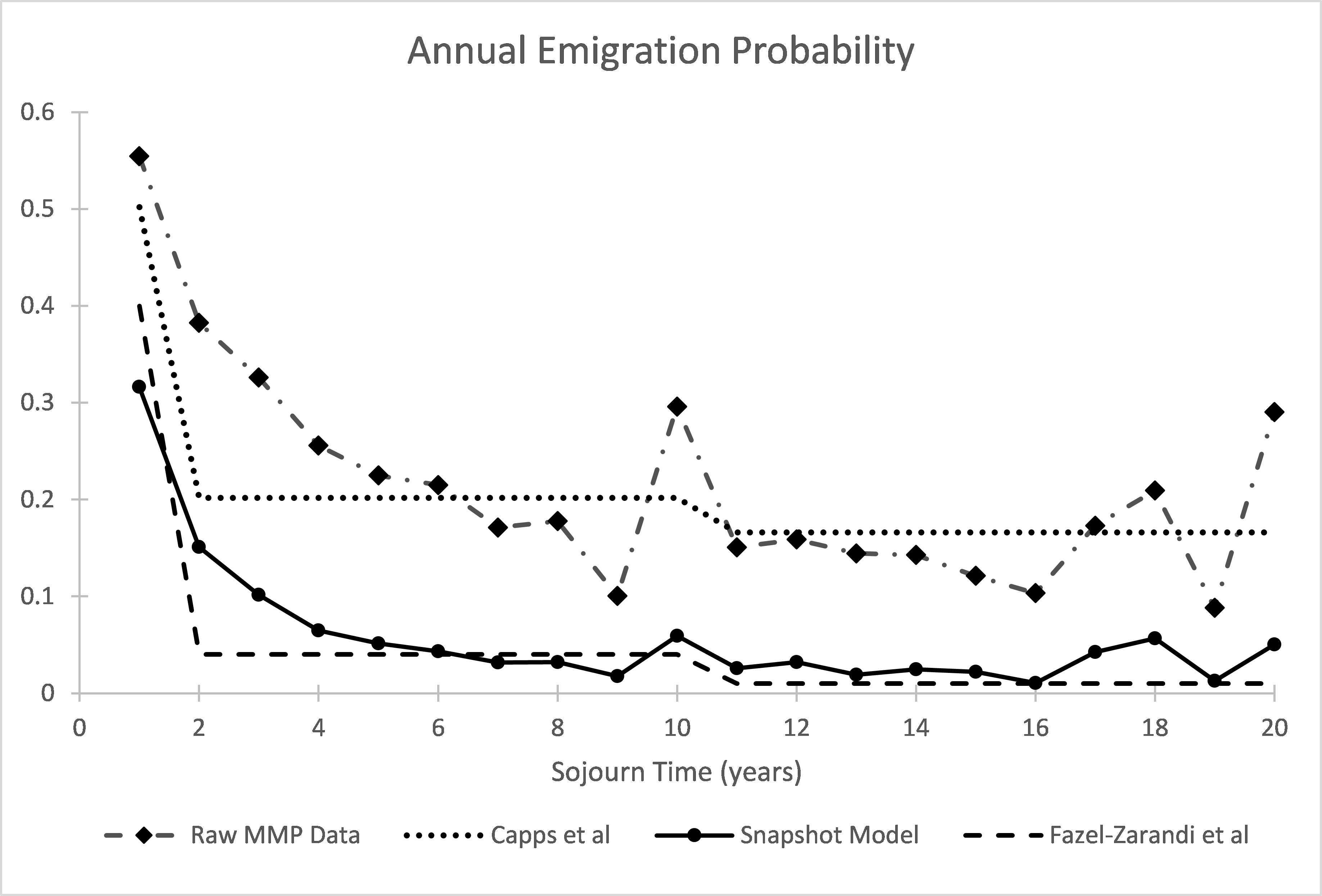}
	\caption{Annual emigration probabilities from the United States}
\end{figure}

Finally, again scaling by southern border crossings from 1990
through 2016, it is possible to translate the relative undocumented
immigrant populations $P_{S}(t)$ and $P_{C}(t)$ into estimates of the actual
number of undocumented immigrants in the United States over this period.
This is achieved by substituting $b_{S}(j)$ for $\phi f_{S}(j)$ in equation (%
\ref{eq16}), $b_{C}(j)$ for $(1-\phi )m_{C}(\tau )f_{C}(j)$ in equation (\ref%
{eq18}), and adding. The results with 95\% confidence intervals obtained via
the delta method (but conditional on $T$) are shown in Figure 7. The
estimated 2016 undocumented population equals 14.6 million, with a 95\%
confidence interval running from 9.4 to 19.8 million. This estimate provides
a lower bound on the true number of undocumented immigrants for several
reasons. First, as noted above, the 38.8 million border crossings used to
scale this model is itself highly conservative, in that it presumed border
apprehension rates much higher than believed in the years 1990-2004. Second,
the sojourn times in this analysis were restricted to at most 37 years for
all migrants. Allowing sojourn times longer than 37 years would result in a
larger undocumented population in the United States. Indeed, multiplying
both the solitary and circular sojourn time probabilities by the same
positive constant less than one would result in an identical data likelihood 
$\mathcal{\mathcal{L}}_{\mathbf{f,a}}$ but with a larger undocumented
population, so in this sense our model is estimating the smallest
undocumented population consistent with the sojourn times observed in the
data. Third, all of the MMP data are for Mexican undocumented migrants, yet
southern border crossers include large numbers of migrants from Central
American countries further south. It is reasonable to presume that the
further a migrant has traveled to reach the United States, the longer such a
migrant is likely to sojourn there. Finally, the undocumented population in
our model is comprised completely of border crossers. \ It does not consider
non-citizens who enter the United States legally, but then become
undocumented immigrants by overstaying their visas. Fazel-Zarandi, Feinstein
and Kaplan (2018) estimated that there were more than 5 million undocumented
visa overstayers in the United States as of 2016.

\begin{figure}[t]
	\includegraphics[width=0.98\columnwidth]{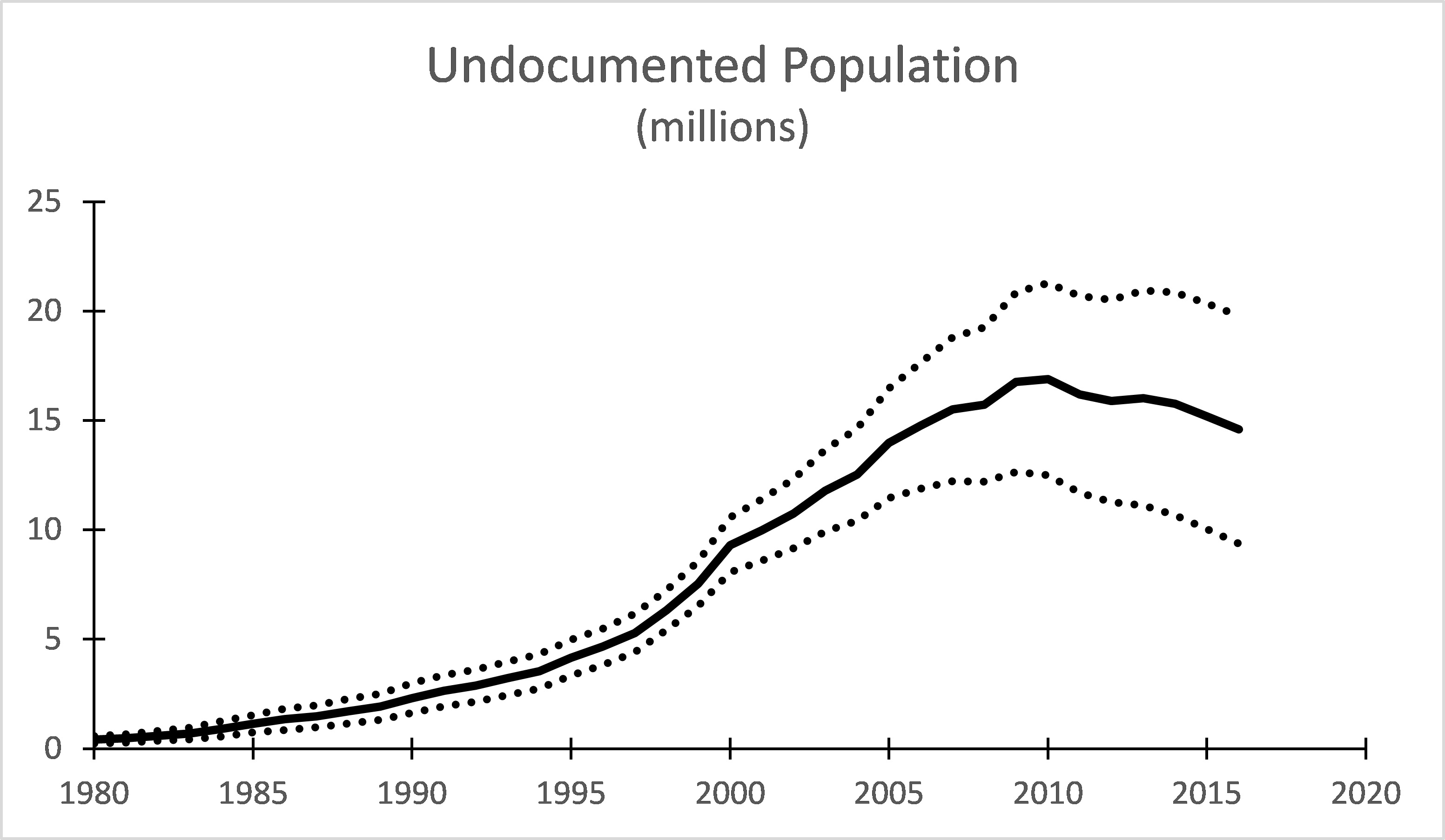}
	\caption{Estimated
		number of undocumented southern border crossers in the United States}
\end{figure}

 The most widely accepted estimates of the number of undocumented
immigrants in the United States hover near 11 million (Pew Research Center,
2018). Fazel-Zarandi, Feinstein and Kaplan (2018) developed estimates based
on demographic modeling that, in 2016, placed a conservative lower bound of
16.7 million persons while via simulation produced a mean estimate of 22
million undocumented immigrants with a 95\% probability interval of 16.2 to
29.5 million. The present model estimates that there are 14.6 million
undocumented border crossers in the United States, with a 95\% confidence
interval running from 9.4 to 19.8 million. Simply adding in the estimated 5
million visa overstayers in the United States would raise our point estimate
to 19.6 million, which is broadly consistent with the results in
Fazel-Zarandi, Feinstein and Kaplan (2018) that were achieved via completely
different methods.

\section{Conclusions}

We have derived a new migration model that enables data-driven estimation of
key quantities such as trip departure and sojourn time probability
distributions, the changing composition of solitary versus circular migrants
over time, and the number of undocumented immigrants in the United States
from snapshot-sampled data with its attendant physical bias. We applied our
model to snapshot-sampled data collected by the Mexican Migration Project,
and in so doing addressed some novel statistical challenges, most notably
mean trip calibration, that enabled appropriate maximum likelihood
estimation and associated uncertainty analysis. While our model does impose
some structural assumptions, the trip departure and sojourn time
distributions estimated are nonparametric, in that no specific probability
distributions were assumed. The numerical results provide migration patterns
that are consistent with prior analyses. In particular, the timing of
circular versus solitary departures and the replacement of the former by the
latter over time is consistent with known security changes along the
southern border.

Regarding the estimated number of undocumented immigrants in the United
States, our model suggests between 9.4 and 19.8 million border crossers in
the US at the end of 2016, with a point estimate of 14.6 million. Our
estimate is larger than the most recent 10.7 million number for \textit{all }%
undocumented immigrants in the US produced via the residual method (Pew
Research Center, 2018). Adding in Fazel-Zarandi, Feinstein and Kaplan's
(2018) estimated 5 million persons who entered the US legally but overstayed
their visas leads to an estimated 19.6 million undocumented immigrants in
total. The estimated population in our new analysis is thus broadly
consistent with Fazel-Zarandi, Feinstein and Kaplan's (2018) earlier study,
even though the data-driven methods of the present paper (and the data
employed) are different. Our work thus adds additional evidence to the
contention that there are many more undocumented immigrants in the United
States than has been appreciated to date.

\if false

\noindent \FRAME{itbpFU}{5.6688in}{3.4826in}{0in}{\Qcb{Figure 1: Model
Overview}}{}{Figure}{\special{language "Scientific Word";type
"GRAPHIC";maintain-aspect-ratio TRUE;display "USEDEF";valid_file "T";width
5.6688in;height 3.4826in;depth 0in;original-width 12.6773in;original-height
7.7712in;cropleft "0";croptop "1";cropright "1";cropbottom "0";filename
'Q5HQY200.wmf';file-properties "XPR";}}

\bigskip

\noindent \FRAME{itbpFU}{5.0315in}{2.7043in}{0in}{\Qcb{Figure 2: Mean Trip
Calibration}}{}{Figure}{\special{language "Scientific Word";type
"GRAPHIC";maintain-aspect-ratio TRUE;display "USEDEF";valid_file "T";width
5.0315in;height 2.7043in;depth 0in;original-width 16.1391in;original-height
8.6525in;cropleft "0";croptop "1";cropright "1";cropbottom "0";filename
'Q5HQY201.bmp';file-properties "XPR";}}

\bigskip

\noindent \FRAME{itbpFU}{5.0496in}{2.9551in}{0in}{\Qcb{Figure 3: Observed
and expected in-sample trip departure date frequency distributions}}{}{Figure%
}{\special{language "Scientific Word";type "GRAPHIC";maintain-aspect-ratio
TRUE;display "USEDEF";valid_file "T";width 5.0496in;height 2.9551in;depth
0in;original-width 6.5691in;original-height 3.8337in;cropleft "0";croptop
"1";cropright "1";cropbottom "0";filename
'Q5HQY202.bmp';file-properties "XPR";}}

\bigskip

\noindent \FRAME{itbpFU}{5.0522in}{2.7371in}{0in}{\Qcb{Figure 4: Observed
and expected in-sample sojourn time frequency distributions}}{}{Figure}{%
\special{language "Scientific Word";type "GRAPHIC";maintain-aspect-ratio
TRUE;display "USEDEF";valid_file "T";width 5.0522in;height 2.7371in;depth
0in;original-width 7.1252in;original-height 3.8475in;cropleft "0";croptop
"1";cropright "1";cropbottom "0";filename
'Q5HQY203.bmp';file-properties "XPR";}}

\bigskip

\noindent \FRAME{itbpFU}{5.0635in}{3.0519in}{0in}{\Qcb{Figure 5: Solitary
versus circular southern border crossings}}{}{Figure}{\special{language
"Scientific Word";type "GRAPHIC";maintain-aspect-ratio TRUE;display
"USEDEF";valid_file "T";width 5.0635in;height 3.0519in;depth
0in;original-width 5.009in;original-height 3.0087in;cropleft "0";croptop
"1";cropright "1";cropbottom "0";filename
'Q5HQY204.wmf';file-properties "XPR";}}

\bigskip

\noindent \FRAME{itbpFU}{4.6959in}{3.1903in}{0in}{\Qcb{\noindent Figure 6:
Annual emigration probabilities from the United States}}{}{Figure}{\special%
{language "Scientific Word";type "GRAPHIC";maintain-aspect-ratio
TRUE;display "USEDEF";valid_file "T";width 4.6959in;height 3.1903in;depth
0in;original-width 6.1211in;original-height 4.1485in;cropleft "0";croptop
"1";cropright "1";cropbottom "0";filename
'Q5HQY205.wmf';file-properties "XPR";}}

\bigskip

\noindent \FRAME{itbpFU}{5.2295in}{3.0519in}{0in}{\Qcb{Figure 7: Estimated
number of undocumented southern border crossers in the United States}}{}{%
Figure}{\special{language "Scientific Word";type
"GRAPHIC";maintain-aspect-ratio TRUE;display "USEDEF";valid_file "T";width
5.2295in;height 3.0519in;depth 0in;original-width 5.1733in;original-height
3.0087in;cropleft "0";croptop "1";cropright "1";cropbottom "0";filename
'Q5HQY206.wmf';file-properties "XPR";}}

\fi

\end{document}